\documentclass[3p,sort&compress]{elsarticle}
\usepackage{graphicx}
\usepackage{hyperref}
\usepackage{color}
\usepackage{bm}
\usepackage{amssymb}
\usepackage{amsmath}
\usepackage[normalem]{ulem}
\usepackage[linesnumbered]{algorithm2e}
\usepackage[utf8]{inputenc} 
\usepackage{multirow}
\usepackage{xcolor}
\usepackage[english]{babel}

\usepackage{verbatim}
\usepackage{float}
\usepackage[section]{placeins}

\begin{document}
\title{
Enhancement of superconductivity by disorder in Remeika-type quasiskutterudites}
\author[1]{Andrzej \'{S}lebarski\corref{cor1}}
\ead{andrzej.slebarski@us.edu.pl}
\author[2]{Maciej M.~Ma\'ska} 
\ead{maciej.maska@pwr.edu.pl}
\cortext[cor1]{Corresponding author}
\affiliation[1]{organization={Institute of Low Temperature and Structure Research, Polish Academy of Sciences},
addressline={Okólna 2}, 
postcode={50-422},
city={Wrocław}, 
country={Poland}}
\affiliation[2]{organization={Institute of Theoretical Physics, Wrocław University of Science and Technology},
addressline={Wybrzeże Stanisława Wyspiańskiego 27}, 
postcode={50-370},
city={Wrocław}, 
country={Poland}}

\begin{abstract}

Atomic-scale disorder is conventionally regarded as detrimental to superconductivity; however, under specific conditions, it can enhance superconducting properties. Here, we investigate the role of substitutional disorder in Remeika-type quasiskutterudites $R_3M_4$Sn$_{13}$ and $R_5M_6$Sn$_{18}$ ($R=$ Y, La, Lu; $M=$ Co, Rh, Ru) by combining measurements of magnetic susceptibility, electrical resistivity, and heat capacity with microscopic modeling. We demonstrate that increasing disorder leads to the emergence of locally superconducting regions characterized by an enhanced critical temperature $T_c^{\ast}$, exceeding the bulk transition temperature $T_c$.

Both $T_c^\ast$ and $T_c$ exhibit a nonmonotonic dependence on dopant concentration and show a strong correlation with entropy isotherms measured as a function of disorder. The pronounced entropy maxima coincide with the largest separation between $T_c^{\ast}$ and $T_c$, establishing disorder as a thermodynamically controlled parameter governing superconductivity in these materials. Measurements of the upper critical field reveal distinct $H_{c2}(T)$ branches associated with the bulk and locally superconducting phases, providing direct experimental evidence for a percolative superconducting state.

To interpret these observations, we propose a microscopic model that captures the interplay between the impurity-induced enhancement of local pairing and the disorder-driven suppression of global superconducting coherence. The model reproduces the experimentally observed nonmonotonic evolution of $T_c^{\ast}$ with disorder and supports a percolation-based interpretation of the superconducting transition. Our results demonstrate that controlled atomic disorder can serve as an effective materials-design parameter for tuning superconductivity in complex correlated systems.


\end{abstract}


\maketitle
\section{Introduction}

Atomic-scale disorder \footnote{Disorder refers to any kind of imperfection or randomness in a material, such as: impurities (foreign atoms), vacancies, lattice defects, and/or local structural inhomogeneities.} can significantly influence the physical properties of solids, particularly at low temperatures, where competing electronic ground states are highly sensitive to even weak perturbations. Strongly correlated electron systems (SCESs) provide prominent examples of this behavior, as disorder decisively modifies local electronic environments and can fundamentally reshape the macroscopic quantum state of a material (cf. \cite{Spalek2003,Mydosh1999,Seo2014}).
Of particular interest are the unconventional forms of quantum magnetism that frequently emerge in disordered SCESs, as well as their coexistence with superconductivity \cite{Steglich1979,Mathur1998,Hegger2000,Petrovic2001a,Petrovic2001b}. Despite the critical role of disorder in the regulation of these phenomena, its direct structural characterization--especially when the disorder is local or mesoscopic in nature--remains challenging with the use of conventional metallurgical techniques. Consequently, the presence and impact of disorder are often inferred indirectly, through the observation of unexpected or anomalous physical behavior.

The role of disorder is especially intriguing in superconductors, where it is traditionally considered detrimental. According to Anderson’s theorem, conventional superconductors \cite{Bardeen1957} are weakly affected by small concentrations of random nonmagnetic impurities  \cite{Anderson1959}, whereas a strong disorder - particularly in the presence of magnetic dopants--leads to pair breaking, suppression of the superconducting transition temperature $T_c$ \cite{Abricosov1961,Kim1993}, and, in extreme cases, to a superconductor–insulator transition \cite{Goldman1998}. However, between these two limits lies a much less explored regime in which disorder can enhance superconductivity (cf. \cite{Gastiasoro2018}). Experimental realizations of disorder-enhanced superconductivity have been reported in a variety of systems (e.g., Ce$_{1-x}$Pr$_x$Ru$_2$, Ce$_{1-x}$Gd$_x$Ru$_2$,~\cite{Matthias1958}), including heavy-fermion compounds (CeIrIn$_5$~\cite{Bianchi0}, CePt$_3$Si~\cite{Kim2005}), filled skutterudites (PrOs$_4$Sb$_{12}$,~\cite{Maple2002,Vollmer2003,Seyfarth2006,McBriarty2009,Meason2008}), cuprates (Bi$_2$Sr$_2$CaCu$_2$O$_{8+\delta}$-type materials \cite{Cren2000,Andersen2006}), layered transition-metal dichalcogenides (TaS$_2$ \cite{Peng2018} and NbSe$_2$, \cite{Shao2019}), and irradiation-disordered superconductors (e.g., Ref. \cite{Leroux2019}). These observations challenge the conventional paradigm and highlight disorder as a potentially constructive ingredient in correlated superconductors.

From a broader perspective, disorder in SCESs is rarely purely random. Structural defects, vacancies, and dopants often exhibit spatial correlations, leading to the emergence of inhomogeneous electronic states. This correlated disorder has been implicated in the formation of {\it unconventional magnetic} behavior \cite{Dobrosavlevic1992,Miranda1996,Miranda1997,Bernal1995,Andraka1993,Seaman1993,Maple1995,Andraka1994a,Andraka1994b}, Griffiths phases   \cite{Griffiths1961,Castro1998,Castro2000,Volmer2000,Slebarski2023,Slebarski2025}, and/or nanoscale phase separation \cite{Slebarski2020c}. In superconductors, it can give rise to spatially modulated superconducting order parameters \cite{Gastiasoro2018}, multistep transitions (cf. \cite{Maple2002}), and superconducting states that persist locally above the
bulk $T_c$ (cf. \cite{Slebarski2020d}). 

The percolating superconductivity in the temperature range above the critical temperature $T_c$ seems to be a good example of such defect correlation behavior \cite{Slebarski2020a,Slebarski2020d}. 
Understanding how these inhomogeneous states evolve and eventually establish global coherence remains a central challenge in the physics of disordered superconductors.
In this context, the family of Remeika-type quasiskutterudites \cite{Remeika1980} provides an exceptionally suitable material platform for investigating disorder-driven superconductivity. Compounds of the form $R_3M_4$Sn$_{13}$ and $R_5M_6$Sn$_{18}$  ($R =$ rare earth or Y and Sr; $M=$ transition metal) exhibit relatively high superconducting transition temperatures, complex cage-like crystal structures \cite{Gumeniuk2018}, and a strong propensity for atomic-scale disorder. Previous studies have documented large static atomic displacements \cite{Slebarski2020a}, local structural inhomogeneities on the scale of superconducting coherence length $\xi$ \cite{Slebarski2020c}, and an unusual sensitivity of superconducting properties to subtle changes in stoichiometry \cite{Deniszczyk2022}, applied pressure or chemical doping \cite{Slebarski2015}.

Our recent investigations of both cubic $R_3M_4$Sn$_{13}$ and tetragonal $R_5M_6$Sn$_{18}$ superconductors have revealed a recurring and robust phenomenon: the appearance of a locally disordered superconducting phase characterized by a higher critical temperature $T_c^{\ast}$, exceeding that of the more ordered bulk phase. The $T_c^{\ast}$-phase manifests itself through broadened superconducting transitions, anomalies in thermodynamic and electrical transport properties above $T_c$, and distinct upper critical field behavior.  These features point to the formation of superconducting islands embedded in a normal matrix, which gradually percolate upon cooling to establish global superconductivity.

The present work aims to elucidate the relationship between atomic disorder and superconductivity in Remeika-type quasiskutterudites in a systematic and
experimentally grounded manner. 
We focus on Ca-doped La$_3$Rh$_4$Sn$_{13}$ and
Y$_5$Rh$_6$Sn$_{18}$, where substitutional disorder can be tuned in a controlled way over a wide concentration range while preserving the single-phase crystal structure. By combining measurements of magnetic susceptibility, electrical resistivity, and heat capacity, we track the
evolution of both the bulk superconducting transition temperature $T_c$ and the higher local transition $T_c^{\ast}$ as a function of disorder.
We used entropy isotherms as a thermodynamic measure of disorder. We show that the evolution of $T_c^{\ast}$ is well correlated with  the change in entropy as a function of dopant concentration, providing direct experimental evidence that disorder--not
simply changes in carrier concentration--controls the enhancement of superconductivity. Furthermore, analysis of the upper critical fields reveals clear signatures of percolative superconductivity, consistent with an inhomogeneous superconducting state.

To support and interpret the experimental findings, we introduced a simplified microscopic model that captures the essential competition between impurity-induced enhancement of the local pairing interaction and the simultaneous suppression of global superconducting
coherence due to disorder-induced changes in the electronic structure. 
Our theoretical modeling reproduces key experimental trends and provides a transparent physical picture of disorder-enhanced superconductivity in these materials.

The paper is organized as follows. In Section 2, we review the current state of knowledge on superconductivity and disorder in Remeika-type compounds. Section 3 describes the experimental procedures. Section 4 presents the experimental results and their analysis,
focusing on the evolution of $T_c$, $T_c^{\ast}$, entropy isotherms, and upper critical fields. In Section 5, we introduce the theoretical model and discuss its implications. Finally, Section 6 summarizes the main conclusions and outlines broader implications for disorder-driven
superconductivity in strongly correlated materials.

\section{Enhancement of superconductivity by disorder in Remeika-phases; Current state of knowledge.\label{sec:enhan}}

Remeika-type quasiskutterudites constitute a broad family of intermetallic compounds with the general formulas $R_3M_4$Sn$_{13}$ \cite{Remeika1980,Espinosa1980,Hodeau1982} and $R_4M_5$Sn$_{18}$ (cf. \cite{Gumeniuk2018} and the references therein included). Since their discovery, these materials have attracted sustained interest because of the strong electron–lattice coupling and pronounced sensitivity to external tuning parameters such as chemical substitution, pressure, and disorder. Several compounds within this family exhibit unconventional superconducting
behavior and/or multiband superconductivity \cite{Kase2011,Juraszek2024}, including proximity to structural quantum critical points \cite{Klintberg2012,Goh2015,Yu2015,Kuo2014,Kuo2015}, and anomalous thermodynamic and transport responses. 
Importantly, even high-quality single crystals frequently display significant atomic-scale disorder in the form of vacancies, site mixing, or large static atomic displacements. These intrinsic imperfections lead to local structural and electronic inhomogeneities on length scales comparable to the superconducting coherence length $\xi$, as previously shown in both cubic $R_3M_4$Sn$_{13}$ \cite{Slebarski2020a,Slebarski2020d}  and tetragonal $R_5M_6$Sn$_{18}$ \cite{Slebarski2020c,Deniszczyk2022,Slebarski2020b,Slebarski2021,Fijalkowski2021,Slebarski2022} systems.
Over the past decade, we have demonstrated experimental evidence that disorder introduced through chemical substitution \cite{Slebarski2015,Fijalkowski2021,Slebarski2022,Slebarski2016}, non-stoichiometry \cite{Slebarski2021}, local inhomogeneity \cite{Slebarski2020c}, or intrinsic lattice imperfections \cite{Slebarski2020a} on the length scale of the coherence length $\xi\sim 10-20$ nm \cite{Slebarski2014} leads to the emergence of a locally superconducting phase characterized by a critical temperature $T_c^{\ast}$ higher than that of the bulk phase.
We have proposed a phenomenological model, which explains the increase in $T_c$ due to the higher stiffness of the locally inhomogeneous {\it high temperature $T_c^{\ast}$-phase} (see, e.g., \cite{Slebarski2016,Slebarski2020d}). 
This approach well explains the relation between the Gr\"{u}neisen parameters, $\Gamma_G$, of the locally disordered superconducting phase $T_c^{\ast}$ and the bulk phase $T_c$ \cite{Slebarski2020d}. That is, $\Gamma_G$ has always been obtained larger for the $T_c^{\ast}$ phase with respect to the bulk one, suggesting a larger stiffening of the lattice for the disordered phase and provides arguments for the relationships $T_c^{\ast}>T_c$ and $\mid \frac{dT_c^{\ast}}{dP}\mid>\mid \frac{dT_c}{dP}\mid$.
Within the inhomogeneous superconductivity scenario, the anomaly at $T_c^{\ast}$ marks the
onset of a locally disordered superconducting phase with a spatial distribution of the magnitude of the superconducting energy gaps $\Delta$. 
It has been found \cite{Slebarski2014} that a simple Gaussian gap distribution $f(\Delta)\varpropto \exp[-\frac{(\Delta-\Delta_0)^2}{2D}]$ provides a good approximation of the anomalous enhancement in specific heat, $\Delta C(T) \equiv C(T,B=0) - C(T,B>H_{c2})$ 
within the temperature range of the broad transition to the superconducting state around $T_c^{\ast}>T_c$. Here, $\Delta_0$ and the variance $D$ of the distribution are fitting parameters. 
In the "dirty limit", characterized by a high concentration of impurities and an electronic mean free path $l\ll\xi$, the mesoscopic size disorder gives rise to a transition width of approximately 1--2 K, this is a case of Ce$_3$(Ru,Co)$_4$Sn$_{13}$ \cite{Slebarski2015}.
For these strongly disordered alloys, both  $C(T)$ and $\chi_{ac}(T)$ exhibit broad maxima near $T_c^{\ast}$, well described by the function $f(\Delta)$, which overlapped with  the bulk transition at $T_c$.
One can suggest that such a broadened superconducting phase transition  may lead to uncertainty in determining the exact critical temperature  $T_c$.

\section{Experimental details\label{sec:exp}}

\begin{figure}[!htb]
\centering
\includegraphics[width=0.7\textwidth]{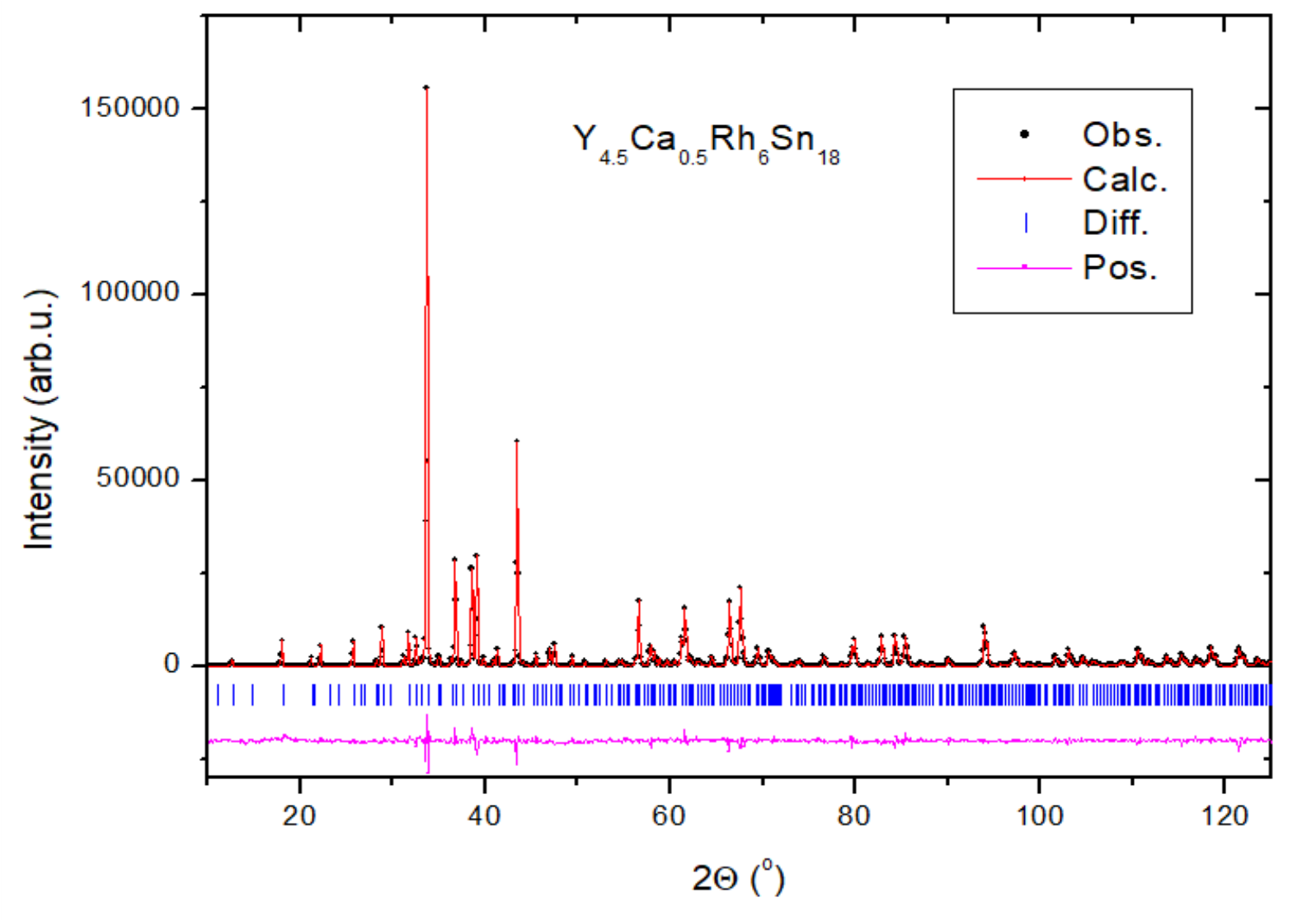}
\caption{\label{fig:XRD}
Plot of Rietveld refinement for Y$_{4.5}$Ca$_{0.5}$Rh$_{6}$Sn$_{18}$ (tetragonal structure $I4_1/acd$, the lattice parameters in (\AA): $a=13.7678(1)$, $c=27.5430(8)$) obtained with the weighted-profile $R$ factor $R_{wp}=1.9$\%. Black dots - observed pattern, red line - calculated, blue ticks - Bragg peaks positions, magenta line - the difference. 
}
\end{figure}

The Remeika phases of La$_3$Rh$_4$Sn$_{13}$ and Y$_5$Rh$_6$Sn$_{18}$ - both doped with Ca - were prepared by the arc melting technique and subsequent annealing  at 800$^{\circ}$C for 2 weeks.
The samples  were examined by x-ray diffraction (XRD) analysis (PANalytical Empyrean diffractometer equipped with a Cu K$\alpha_{1,2}$ source) and found to be a single-phase polycrystals with a structure respectively: cubic $Pm\bar{3}m$ (La$_{3-x}$Ca$_x$Rh$_4$Sn$_{13}$, $0\leq x \leq 3$), and tetragonal $I4_1/acd$ (Y$_{5-x}$Ca$_x$Rh$_6$Sn$_{18}$, $x \leq 1.3$).
The respective XRD patterns were analyzed with the Rietveld refinement method using the Fullprof Suite set of programs \cite{Rodriguez1993}. Details about the crystal structure of these compounds have previously been published \cite{Slebarski2020a,Slebarski2016,Slebarski2020c,Slebarski2020b}. Figure \ref{fig:XRD} shows, as an example, the XRD pattern of Y$_5$Rh$_6$Sn$_{18}$ doped with Ca, not yet published.
The stoichiometry and homogeneity of the samples were checked using an electron energy dispersive spectroscopy (EDS) technique. 
The atomic percentage of the specific element content in the obtained samples is very close to the assumed composition (it deviates from the assumed composition at an acceptable level).

Magnetic measurements of dc were carried out in the temperature interval $1.8-400$ K employing a superconducting quantum interference device (SQUID) magnetometer.
The magnetic susceptibility of ac, $\chi_{ac}=\chi'+i\chi''$, was measured in the temperature range of 2 to 300 K using a Quantum Design Physical Property Measurement System (PPMS). The measurements were performed under an ac magnetic field of 2 Oe at various excitation frequencies. The maxima in derivative $\frac{d\chi'}{dT}$ and $\frac{d\chi''}{dT}$ we assigned, respectively, to the critical temperatures $T_c^{\ast}$ and $T_c$ \cite{Slebarski2020a}.

The heat capacity, $C$, was measured in the temperature range $0.5-300$ K and in external magnetic fields of up to 9 T using the same PPMS platform.

The electrical resistivity, $\rho$, was investigated using a conventional four-point ac technique using a Quantum Design PPMS. The resistivity measurements were used to determine the critical temperature $T_c^{\ast}$, defined as the temperature at which the resistivity drops to 50 \% of its normal state value. This value of $T_c^{\ast}$ is consistent with that obtained from the measurements of $\chi_{ac}$, cf. \cite{Slebarski2020a}.

Details of the temperature dependencies of $\chi_{ac}(T)$, $C(T)$, and $\rho(T)$ have been previously published (see references listed in the Introduction). In this work, we analyze the changes in $T_c$ and $T_c^{\ast}$ induced by the disorder resulting from doping of the parent superconductors.

\section{Structurally disordered Remeika phases with superconductivity\label{sec:struct}}
\subsection{Enhancing superconductivity by disorder}

\begin{figure}[!htb]
\centering
\includegraphics[width=0.45\textwidth]{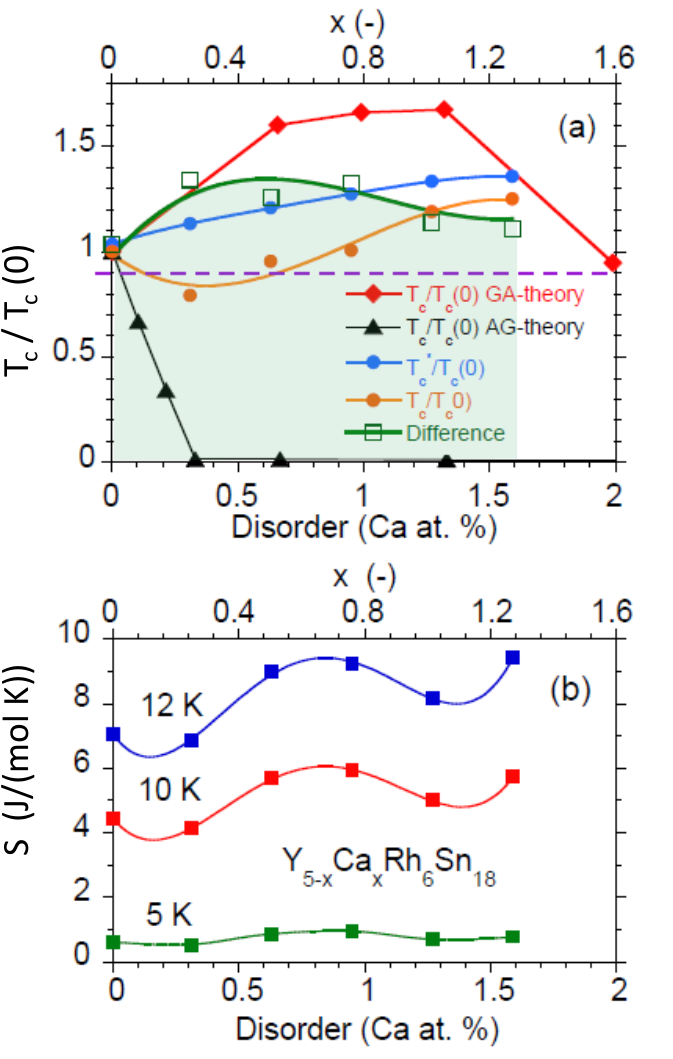}
\caption{\label{fig:Y-Ca_T-x}
$T-x$ diagram for the series of Y$_{5-x}$Ca$_x$Rh$_6$Sn$_{18}$ compounds with tetragonal  structure $I4_1/acd$. (a) Superconducting critical temperatures $T_c$ and $T_c^{\ast}$  are divided by $T_c(x=0)\equiv T_c(0)$ for Y$_5$Rh$_6$Sn$_{18}$.
The blue points represent the locally disordered $T_c^{\ast}$  phase, the orange points
represent the bulk $T_c$ phase, and the green open squares show 
$\frac{1}{T_c(0)}(T_c^{\ast}-T_c)+1$. The solid lines are the best approximations by third degree polynomial. The red rhombic points show disorder-driven enhancement in $T_c$ calculated by a multiband model in an unconventional multiband $s_{\pm}$ SC vs. disorder concentration in atomic \% (the data taken from Ref. \cite{Gastiasoro2018}), while the black triangles show the $T-x$ change consistent with Abrikosov and Gor’kov theory \cite{Abricosov1961} (the data taken from Ref. \cite{Gastiasoro2018}). The respective points are connected by the lines. (b) Entropy isotherms $S$ at $T=5$, 10 and 12 K as a function of Ca at. \% (and $x$).
}
\end{figure}

Experimental evidence for enhancement of superconductivity by disorder remains relatively scarce. 
Various compounds related to ancient compounds \cite{Gumeniuk2018}, including cubic La$_3M_4$Sn$_{13}$ (structure $Pm\bar{3}n$; $M =$ Co, Rh, Ru \cite{Slebarski2014,Mishra2011,Slebarski2015}) 
and tetragonal $R_5$Rh$_6$Sn$_{18}$ (structure $I4_1/acd$; $R =$ Sc, Y, Lu \cite{Deniszczyk2022,Slebarski2020c,Slebarski2021}), all of which exhibit pronounced static atomic displacements \cite{Slebarski2020a} and local structural inhomogeneities on the scale of superconducting coherence length  \cite{Slebarski2014}.
This phenomenon is particularly well documented in the Ca-doped series of Y$_5$Rh$_6$Sn$_{18}$  and La$_3$Rh$_4$Sn$_{13}$. As shown in Fig. \ref{fig:Y-Ca_T-x} (a), moderate Ca substitution results in a substantial enhancement of superconductivity, with $T_c^{\ast}$ increasing by up to approximately 25\% relative to the undoped compound, while the bulk $T_c$ exhibits a weaker and nonmonotonic
response. This intriguing phenomenon can be qualitatively understood  within the Gastiasoro and Andersen (GA) approach \cite{Gastiasoro2018} in a dilute disorder scenario (the case of multiband superconductors). 
Figure \ref{fig:Y-Ca_T-x} (a) displays the enhancement generated by random disorder in $T_c$ for  Y$_{5-x}$Ca$_x$Rh$_6$Sn$_{18}$, when the dopant content is less than the maximum concentration limit of Ca $x = 1.25\equiv 1.6$ at. \% for single-phase samples with tetragonal structure $I4_1/acd$, and its relation to the GA theory \footnote{Y$_5$Rh$_6$Sn$_{18}$ doped with Ca shows a substantial increase in $T_c$ up to 25\% and exhibits emergence of high-$T_c$ disordered superconducting phase with a critical temperature $T_c^{\ast}$ higher than $T_c$ of the bulk sample. See Fig. \ref{fig:Y-Ca_T-x} and Refs.~\cite{Gastiasoro2018,Juraszek2024}.}. 

\begin{figure}[!htb]
\centering
\includegraphics[width=0.65\textwidth]{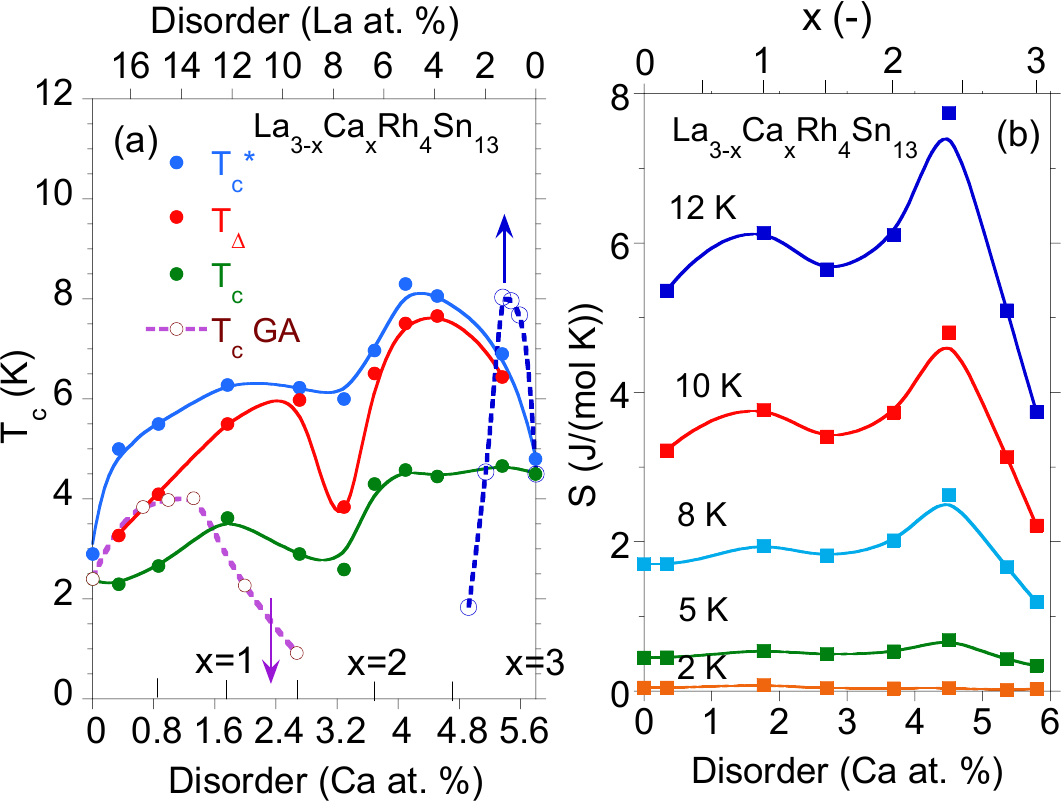}
\caption{\label{fig:La-Ca_disorder}
(a) $T-x$ diagram for the series of La$_{3-x}$Ca$_x$Rh$_4$Sn$_{13}$ compounds. The blue points indicate the beginning of the transition between {\it normal} and SC phase at $T_c^{\ast}$. The red points represent the temperature $T_{\Delta}$ of the maxima of the $f(\Delta)$ function for each component of the series. The green points show the {\it bulk}  $T_c$ phase. The purple and dark blue  dotted curves show disorder-driven enhancement in $T_c$ calculated by a multiband GA model for Ca dopants in La$_{3-x}$Ca$_x$Rh$_4$Sn$_{13}$ (lower at. \% scale) and La  dopants in Ca$_{3-x}$La$_x$Rh$_4$Sn$_{13}$ (upper at. \% scale), respectively (the data taken from Ref. \cite{Gastiasoro2018}). (b) La$_{3-x}$Ca$_x$Rh$_4$Sn$_{13}$, entropy isotherms $S_T$ as a function of Ca dopant in at. \% (and $x$).
}
\end{figure}

Similarly, the $T-x$ diagrams shown in Fig. \ref{fig:La-Ca_disorder}(a) exhibit $T_c$ and $T_c^{\ast}$ vs. $x$ for La$_{3-x}$Ca$_x$Rh$_4$Sn$_{18}$ compounds, all cubic with symmetry $Pm\bar{3}n$. 
The degree of lattice disorder can be expressed here either by increasing the atomic concentration of Ca dopants in La$_{3}$Rh$_4$Sn$_{18}$ or increasing the impurities of La in  Ca$_{3}$Rh$_4$Sn$_{18}$ (data are directed at the upper $x$ axis). 
Within this presentation of the $T-x$-data, the GA theory qualitatively predicts a minimum value of $T_c$ at $x\sim 2$, however, extremally large enhanced in $T_c^{\ast}$, even of $\sim 100$\% with respect to the bulk transition temperature $T_c$, can not be well approximated by this theoretical model.
As shown in Figs. \ref{fig:Y-Ca_T-x}(a) and \ref{fig:La-Ca_disorder}(a),  the GA model roughly predicts the increase in $T_c$ with doping, however, it does not provide quantitative relationships in the behavior of $T_c^{\ast}(x)$.  

A central experimental result of the present work is the close correspondence between the evolution of superconducting transition temperatures and entropy isotherms. Figure \ref{fig:Y-Ca_T-x} (b)
displays the entropy $S(x)$ measured at fixed temperatures as a function of the Ca concentration
for Y$_{5-x}$Ca$_x$Rh$_6$Sn$_{18}$. The entropy exhibits pronounced maxima at intermediate concentrations of dopants, coinciding with the largest separation between $T_c^{\ast}$ and $T_c$. This
correlation provides strong thermodynamic evidence that atomic-scale disorder governs the stabilization of the locally superconducting phase. A similar behavior is observed in the cubic La$_{3-x}$Ca$_x$Rh$_4$Sn$_{18}$ system. The entropy isotherms presented in Fig. \ref{fig:La-Ca_disorder}(b) again display maxima that closely
follow the evolution of $T_c^{\ast}$, reinforcing the conclusion that disorder--rather than simple band filling effects--controls the observed enhancement of superconductivity.
Although theoretical approaches based on diluted disorder in multiband superconductors qualitatively reproduce the increase of $T_c$ with disorder, they do not fully account for the experimentally observed behavior of $T_c^{\ast}$. The present results indicate that locally disordered superconducting regions play a crucial role and that their spatial distribution must be taken into account to describe the superconducting transition in these systems.

\subsection{Temperature dependencies of the upper critical fields $H_{c2}$ in the $H-T$ diagrams}

\begin{figure}[!htb]
\centering
\includegraphics[width=0.45\textwidth]{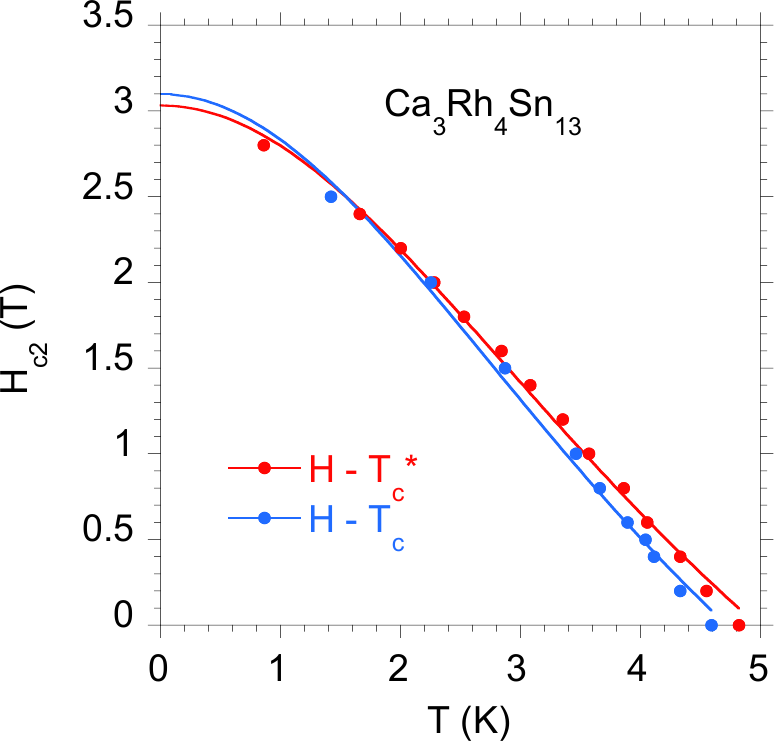}
\caption{\label{fig:Fig_Ca3Rh4Sn13_H-T}
The upper critical field $H_{c2}$ vs. $T^\ast$ for Ca$_{3}$Rh$_4$Sn$_{13}$. The $H-T$ data are approximated by GL equation $H_{c2}(T)=H_{c2}(0)\frac{1-t^2}{1+t^2}$, where $t=\frac{T}{T_c}$. 
} 
\end{figure}

Further insight into the nature of the superconducting state is obtained from the temperature dependence of the upper critical field $H_{c2}$. 
The percolation superconductivity for systems with local disorder
assumes different mesoscopic regions in the sample volume, which may slightly differ in composition depending on the amount of defects, and in result determines the various relationships between the macroscopic values of $H_{c2}$ and $T_c$ or $T_c^{\ast}$ of these material. Ca$_3$Rh$_4$Sn$_{13}$ with experimentally documented disorder \cite{Westerveld1987,Westerveld1989,Slebarski2016} fits well this scenario and shows two different branches $H_{c2}(T)$, where one is related to the bulk phase and the other to the $T_c^{\ast}$ phase, as shown in Fig. \ref{fig:Fig_Ca3Rh4Sn13_H-T}.
\begin{figure}[!htb]
\centering
\includegraphics[width=0.48\textwidth]{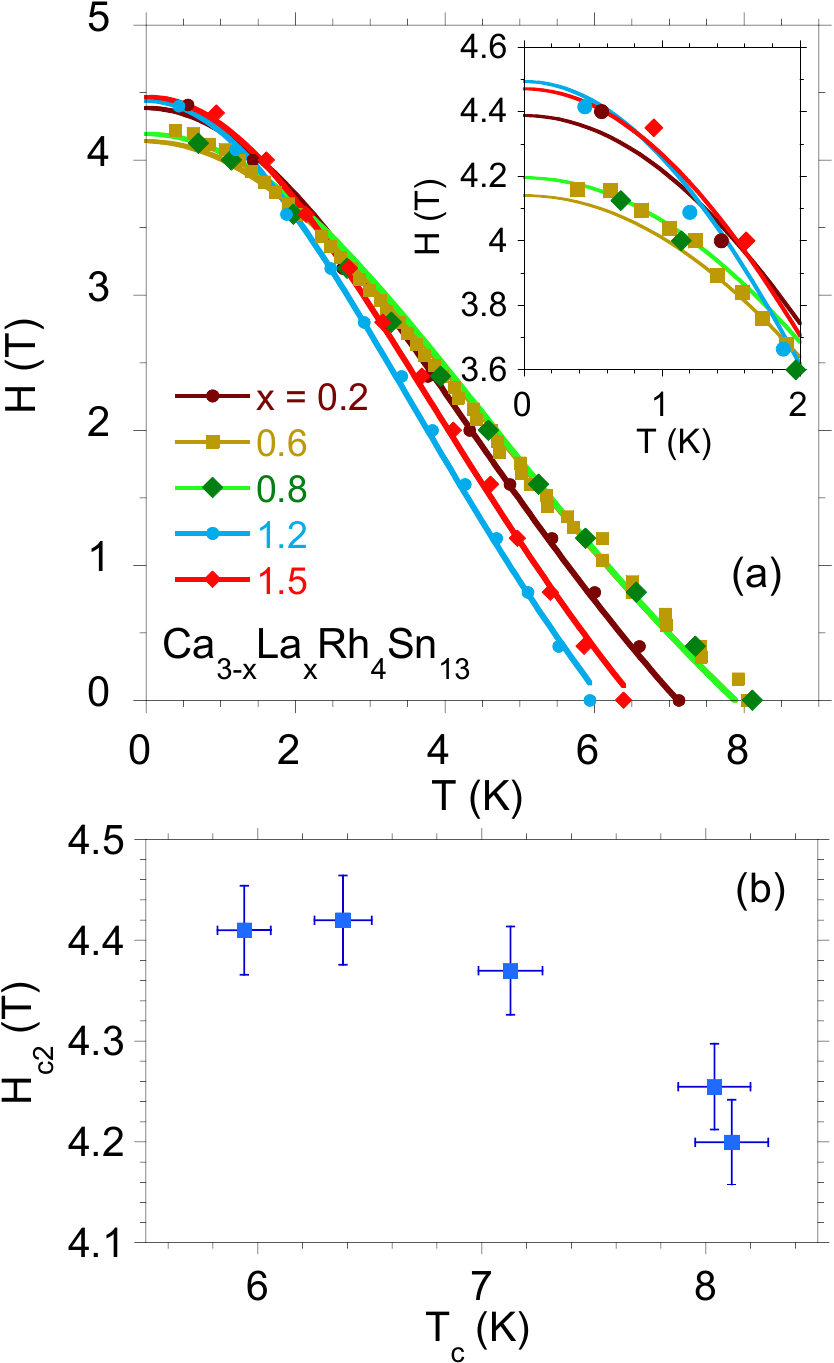}
\caption{\label{fig:La-Ca_H-T}
The upper critical field $H_{c2}^{\ast}$ vs. $T$ for Ca$_{3}$Rh$_4$Sn$_{13}$ doped with La. (a) Ca$_{3-x}$La$_x$Rh$_4$Sn$_{13}$; the $H-T$ data are approximated by GL equation $H_{c2}(T)=H_{c2}(0)\frac{1-t^2}{1+t^2}$, where $t=\frac{T}{T_c}$. (b) $H_{c2}^{\ast}$ at $T=0.4$ K as a function of $T_c^{\ast}$ at $H=0$ for the series of Ca$_{3-x}$La$_x$Rh$_4$Sn$_{13}$ superconductors with various level of disorder due to doping.
} 
\end{figure}
The coexistence of these branches provides direct experimental
evidence for spatially inhomogeneous superconductivity.
An even more pronounced manifestation of this behavior is found in La-doped Ce$_3$Rh$_4$Sn$_{13}$, where disorder is further enhanced by chemical substitution.  As illustrated in Fig. \ref{fig:La-Ca_H-T}(a),  the $H_{c2}$ curves evolve systematically with increasing disorder, while Fig. \ref{fig:La-Ca_H-T}(b) demonstrates a clear correlation between $H_{c2}^{\star}$ and $T_c^{\star}$. Notably, the enhancement of $T_c^{\star}$ is accompanied by a reduction of the corresponding upper critical field, consistent with a percolative superconductivity scenario.
Figure  \ref{fig:La-Ca_H-T}(a) also exhibits  positive curvature in $H_{c2}^{\star}$ below $T_c^{\star}$ \footnote{This  $H-T$  behavior can be well approximated by the random resistor network (RRN) model \cite{Slebarski2020d} (see also \cite{Zeimetz2002,Bucheli2013}).}, a hallmark of superconductivity emerging from disconnected superconducting regions that progressively establish long-range coherence upon cooling. 
The $H_{c2}^{\star}$ data provide compelling support for a percolative model in which superconductivity in disordered Remeika phases is governed by percolation of locally superconducting regions rather than a uniform bulk instability.

\section{A simple model for $T_c^*$ in disordered superconductors\label{sec:model}}
Here we introduce a microscopic toy model that accounts for the nonmonotonic dependence of the superconducting critical temperature $T_c^*$ on the strength of the disorder, as experimentally observed in Figs.~\ref{fig:Y-Ca_T-x}(a) and \ref{fig:La-Ca_disorder}(a). The nonmonotonic behavior arises from the competition between two disorder-induced effects. The first is a local enhancement of the pairing interaction in the vicinity of impurities. The second is the suppression of global superconducting coherence due to the breakdown of percolating superconducting paths caused by increasing disorder.

The model is based on two key assumptions. ({\it i}\,) The strong disorder leads to a highly inhomogeneous superconducting state consisting of coexisting superconducting and normal regions. ({\it ii}\,) Within the superconducting regions, the pairing interaction is locally enhanced by an inhomogeneous atomic potential induced by impurities. The latter effect has been observed and discussed in a variety of experimental and theoretical studies, including works by the present authors
\cite{Maska2007,Mashima2006,McElroy2005,Hirschfeld2009,Hirschfeld2010,Khaliullin2010,Hirschfeld2012,Goshal2017}. While several microscopic mechanisms have been proposed to explain this phenomenon--some of them introduced to account for the correlation between the magnitude of the superconducting gap and the position of oxygen dopants in cuprates \cite{Maska2007}--here we adopt a phenomenological approach without committing to a specific microscopic origin.


Assumption ({\it i}\,) implies that the superconducting transition is of a percolative nature. The spatial distribution of superconducting and normal regions depends on temperature and magnetic field. As the temperature is reduced but remains above the global critical temperature $T_c^*$, the volume fraction occupied by disconnected superconducting regions increases. Within the framework of the random resistor network (RRN) model, this corresponds to an increasing number of links that acquire zero resistance, leading to a gradual decrease in the resistivity of the sample \cite{Slebarski2020d}. At sufficiently low temperature, superconducting regions form a continuous percolation path across the system, resulting in zero resistivity and defining the superconducting transition.

To implement assumption ({\it ii}\,), we introduce a position-dependent pairing interaction $J(\bm r)$ whose strength increases in the vicinity of an impurity,
\begin{equation}
    J({\bm r})=J_0+J_1\sum_i f({\bm r},{\bm r}_i),
\end{equation}
where $J_0$ is the homogeneous, position-independent contribution and $J_1f(\bm r,\bm r_i)$ describes the modification of the pairing interaction induced by the $i^{\mathrm{th}}$ impurity located at $\bm r_i$. Although the summation formally runs over all impurities, the short-range nature of the screened impurity potential allows it to be restricted to a finite radius $r_a$ around $\bm r$. While the precise form of $f(\bm r,\bm r_i)$ does not affect the qualitative behavior of the model, it is assumed to be positive in the vicinity of an impurity.

In this work, we consider a two-dimensional square lattice and assume that the impurity-induced contribution to the pairing interaction is proportional to the squared magnitude of the gradient of the screened impurity potential,
\begin{equation}
    f({\bm r})=\sum_i\sum_{\bm \delta}\left[V({\bm r},{\bm r}_i)-V({\bm r}+{\bm\delta},{\bm r}_i)\right]^2,
    \label{eq:pairing_correction}
\end{equation}
where $\bm r_i$ is the position of $i^{\rm th}$ impurity and $\bm \delta\in\{+a\hat{\bm x},-a\hat{\bm x},+a\hat{\bm y},-a\hat{\bm y}\}$, $a$ is the lattice constant, and $\hat{\bm x}$ and $\hat{\bm y}$ are unit vectors along the lattice directions. The parameter $J_1$ controls the relative strength of the impurity-induced contribution to the pairing interaction. For pairing mediated by Coulomb-interaction-induced superexchange, this form emerges as the leading-order correction \cite{Maska2007}, with $A$ inversely proportional to the square of the Coulomb interaction. For other pairing mechanisms, the leading correction may be linear rather than quadratic in the potential gradient; however, such a modification would affect only quantitative aspects of the results.

The impurity potential is modeled by a Yukawa-type screened interaction,
\begin{equation}
V_i({\bm r})\propto \frac{\exp(-r/r_0)}{r},
\label{eq:yukawa}
\end{equation}
where $r=|\bm r-\bm r_i|$ and $r_0$ defines the screening length. We include the coefficient in Eq.~\eqref{eq:yukawa} in the definition of $J_1$.

\begin{figure}[!h]
\centering
\includegraphics[width=0.7\textwidth]{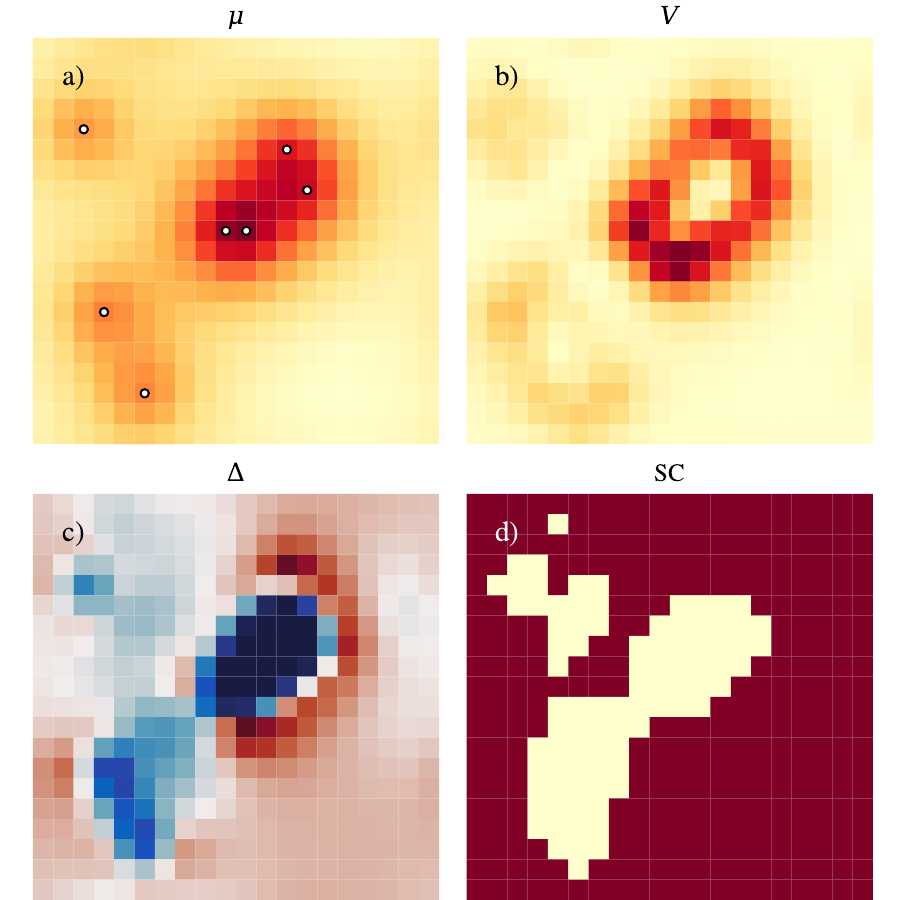}
\caption{a) Increase of the atomic level close to impurities (represented by white circles), b) increase of the pairing potential, c) impurity-induced change of the pairing amplitude: red represents an increase and blue represents a decrease, d) dark color represent regions, where the pairing amplitude is larger than the threshold value $\Delta_0$, which we assume to be superconducting.
\label{fig:single_disorder}}
\end{figure}

To study the influence of impurity strength and concentration on the superconducting transition, we fix the model parameters, temperature, and impurity positions, and solve the Bogoliubov--de Gennes (BdG) equations to obtain the spatial distribution of the pairing amplitude $\Delta(\bm r)$. We assume that regions that satisfy $|\Delta(\bm r)|>\Delta_0$, where $\Delta_0$ is a small threshold value, are locally superconducting. An example of this procedure for a representative impurity configuration is shown in Fig.~\ref{fig:single_disorder}: panel (a) displays the impurity-modified chemical potential, panel (b) the resulting enhancement of the pairing interaction, panel (c) the solution for the pairing amplitude, and panel (d) the corresponding superconducting regions.

Within the RRN framework \cite{Slebarski2020d}, an increase in the total area of superconducting regions leads to a reduction of the sample resistivity. As the temperature decreases, superconducting regions expand and may eventually form a percolating path across the system. We define the corresponding temperature as the critical temperature $T_c^{(i)}$ for the $i^{\mathrm{th}}$ disorder realization. Repeating this procedure for $M$ independent disorder realizations allows us to determine the probability that a given temperature $T$ lies below the critical temperature,
\begin{equation}
    p(T) \equiv P(T<T_c^*) = \frac{1}{M}\sum_{i=1}^M\Theta(T_c^{*(i)}-T),
\end{equation}
where $\Theta(\ldots)$ denotes the Heaviside step function. To model $p(T)$, we use a supervised machine learning technique known as binomial logistic regression. In this approach, the probability of occurrence of a dichotomous event (i.e., whether or not a percolation path exists) is predicted by fitting independent variables to a logistic curve
\begin{equation}
    p(T)=\frac{1}{1+e^{(T-T_c^*)/s}},
    \label{eq:logistic}
\end{equation}
where $T_c^*$ and $s$ are fitting parameters. The fitting was performed using the standard technique of maximizing the likelihood function \cite{mccullagh1989glm}.
\begin{figure}[!htb]
\centering
\includegraphics[width=0.4\textwidth]{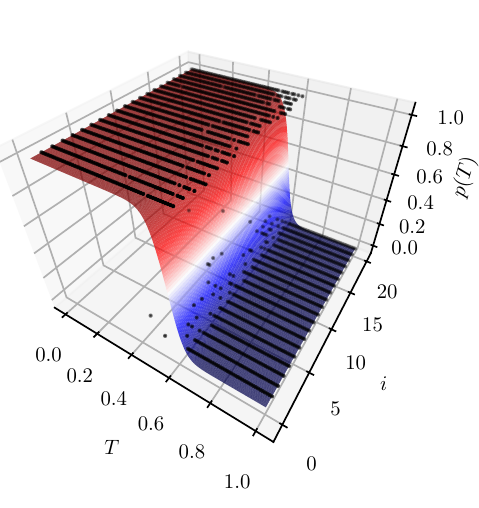}
 \caption{Fit of the logistic curve $p(T)$ to the results of the simulation for 5 impurities in a $20\times 20$ system. $i\in [1,\ldots,20]$ enumerates the random configurations of the impurities; black points at $p(T)=1$ [at $p(T)=0$] represent simulations for which the percolation path existed [did not existed]. The fitted logistic curve $p(T)$ is represented by the cross section of the S-shaped surface with the critical temperature given by $p(T^\ast_c)=0.5$. 
 \label{fig:logistic}}
\end{figure}
Figure \ref{fig:logistic} illustrates the logistic curve fitted to the percolation results in the case of 5 impurities in a $20\times 20$ system. Different rows of data points (0 or 1 for the absence or presence of the percolation path, respectively) describe different disorder realizations.

For each set of model parameters, we generated 500 disorder realizations. We solved the BdG equations for each realization and determined whether a percolation path existed. Using these results, we determined $T_c^*$ by fitting the logistic curve \eqref{eq:logistic}. This process is outlined in Algorithm ~\ref{alg:alg}.

\IncMargin{1em}
\begin{algorithm}[!htb]
\SetKwData{Left}{left}\SetKwData{This}{this}\SetKwData{Up}{up}
\SetKwFunction{Union}{Union}\SetKwFunction{FindCompress}{FindCompress}
\SetKwInOut{Input}{input}\SetKwInOut{Output}{output}
\For{number of impurities $M\leftarrow$ $0$ \KwTo $M_{\max}$}{\label{forimp}
    \BlankLine
    \For{temperature $T\leftarrow$ $0$ \KwTo $T_{\max}$}{\label{fortemp}
        \BlankLine
        $r = 0$\;
        \BlankLine
        \For{disorder realisation $\leftarrow 1$ \KwTo $N$}{
            \BlankLine
            randomly distribute $M$ impurities\;            
            calculate the total potential of the impurities $\Rightarrow\ V({\bm r})$\;
            solve the BdG equations $\Rightarrow\ \Delta({\bm r})$\; 
            determine the superconducting regions as ${\cal D}=\{{\bm r}_i :\ \Delta({\bm r}_i) > \Delta_0$\}\; 
            \BlankLine
            \If{$\mathrm{percolation\ path\ exists\ in\ }{\cal D}$}{
                $r = r+1$\;
            }
            \BlankLine
        }
            \BlankLine
        $P(T,M) = r / N$\;
            \BlankLine
        }
        fit $p(T)$ to $P(T,M)\ \Rightarrow\ T^\ast_c(M)$
        \BlankLine
}
\BlankLine
\caption{The algorithm that was used to calculate the percolative transition temperature $T^\ast_c$ as a function of the number of impurities $M$.\label{alg:alg}}
\end{algorithm}
\DecMargin{1em} 

\begin{figure}[!htb]
\centering
\includegraphics[width=0.8\textwidth]{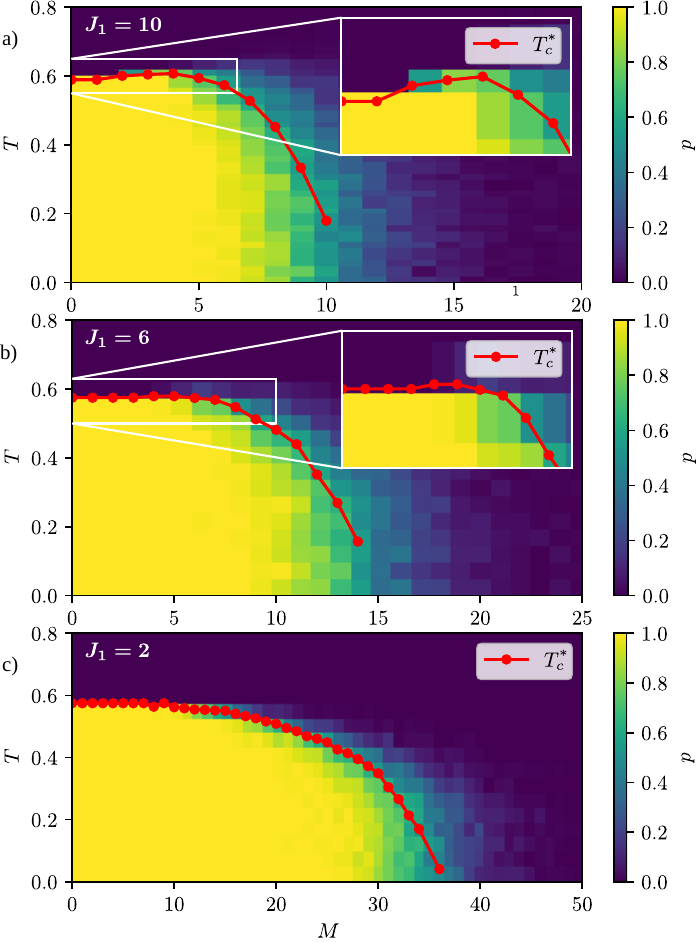}
\caption{The false-color maps show the probability $p$ that a percolation path exists across the system at a given temperature $T$ and for a given number of impurities $M$. Panels a), b), and c) correspond to $J_1=10,\:6$, and $2$, respectively. The red lines show the percolative critical temperature $T^\ast_c$ defined by the condition $p(T^\ast_c,M)=0.5$.
\label{fig:disorder_panels}}
\end{figure}

The results of the toy model are summarized in Fig. \ref{fig:disorder_panels}(a–c), which shows the dependence of the superconducting critical temperature on the number of impurities for different strengths of the impurity potential $A$. The behavior of $T_c^*$ is governed by two competing impurity-induced effects. First, each impurity locally enhances the pairing interaction, leading to an increase of the superconducting order parameter in its vicinity. Second, the impurity potential simultaneously decreases the local chemical potential, effectively reducing the electron concentration and suppressing superconductivity. The global critical temperature is determined by the balance between these two mechanisms in an inhomogeneous, percolative superconducting state.

For strong impurity potentials [Fig. 8(a)], the enhancement of the pairing interaction dominates at low impurity concentrations. As a result, increasing the number of impurities initially leads to an expansion of superconducting regions and a higher probability of forming percolating superconducting paths, giving rise to an increase of the global critical temperature. Upon further increasing the concentration of impurities, the reduction in carrier density and the growing fragmentation of superconducting regions suppress global coherence, and $T_c^*$ decreases.

For the intermediate impurity strength [Fig. 8(b)], the two effects nearly compensate each other. The enhancement of the pairing interaction is weaker, and the initial increase in $T_c^*$ is strongly reduced, resulting in an almost monotonic suppression of the critical temperature with increasing disorder. Finally, for weak impurity potentials [Fig. 8(c)], the pairing enhancement is insufficient to overcome the suppression caused by the reduced electron concentration and the disrupted percolation paths, and $T_c^*$ decreases monotonically with impurity concentration.

Overall, the model demonstrates that increasing the impurity concentration can increase the superconducting critical temperature for a sufficiently strong impurity potential. This provides a microscopic, physically transparent explanation for the behavior observed in our experiments and shown in Figs.~\ref{fig:Y-Ca_T-x}(a) and \ref{fig:La-Ca_disorder}(a).

\section{Summary and conclusions\label{sec:concl}}
In this work, we have demonstrated that atomic-scale disorder can enhance superconductivity in Remeika-type quasiskutterudites, despite the conventional expectation that disorder suppresses superconducting order. By combining systematic experimental studies of Ca-doped La$_3$Rh$_4$Sn$_{13}$ and Y$_5$Rh$_6$Sn$_{13}$ with a phenomenological and microscopic theoretical analysis, we established a coherent picture of disorder-driven superconductivity in these strongly correlated three-dimensional systems.

Experimentally, substitutional disorder leads to the emergence of locally superconducting regions characterized by an enhanced critical temperature $T^\ast_c > T_c$, preceding the bulk superconducting transition. The evolution of both $T^\ast_c$ and $T^\ast_c$ with dopant concentration is nonmonotonic and correlates remarkably well with the entropy isotherms $S(x)|_{T={\rm const}}$, which serve as a thermodynamic proxy for the degree of atomic disorder. These results demonstrate that disorder not only modifies local electronic properties but also governs the macroscopic superconducting response via spatial inhomogeneity on the scale of the coherence length.

The superconducting transition in the presence of strong disorder is shown to be intrinsically percolative. This interpretation is supported by the coexistence of multiple $H_{c2}(T)$ branches associated with the bulk and locally disordered superconducting phases, as well as by the characteristic positive curvature of $H_{c2}(T)$ near $T^\ast_c$, which can be naturally explained within a random resistor network framework.

To explain the observed nonmonotonic dependence of $T^\ast_c$ on disorder, we introduced a microscopic toy model that explicitly incorporates two competing impurity-induced effects: ({\it i}\,) a local enhancement of the pairing interaction in the vicinity of impurities and ({\it ii}\,) a simultaneous suppression of superconductivity due to the effective reduction of the local chemical potential and carrier concentration. Solving the BdG equations for many realizations of disorder and identifying the percolative transition through logistic regression, we showed that the balance between these two mechanisms leads to an increase in $T^\ast_c$ at low impurity concentrations for sufficiently strong impurity potentials, followed by suppression at higher levels of disorder. The resulting phase diagrams reproduce the general trends observed in experiments and present a microscopic explanation for disorder-enhanced superconductivity. While we do not claim that this is the only possible explanation, we offer it as a potential option.

Overall, our results establish atomic disorder as an active tuning parameter for superconductivity in Remeika-type quasiskutterudites. Rather than acting solely as a pair-breaking perturbation, disorder can locally strengthen pairing and promote superconductivity through a percolative mechanism. These findings place quasiskutterudite superconductors among a broader class of strongly correlated materials where controlled disorder offers a viable route toward enhancing superconducting properties and engineering novel inhomogeneous superconducting states.

\section*{CRediT authorship contribution statement} 
Andrzej \'{S}lebarski: Writing – review \& editing, Conceptualization, Visualization, Formal analysis, Methodology, Writing - original draft, Investigation, Software, Data curation. Maciej M. Ma\'{s}ka: Writing – review \& editing, Conceptualization, Visualization, Formal analysis, Methodology, Writing – original draft, Investigation, Software, Data curation.

\section*{Declaration of competing interest} 
The authors declare that they have no known competing financial interests or personal relationships that could have appeared to influence the work reported in this paper.

\section*{Acknowledgement}
Numerical calculations have been carried out using High Performance Computing resources provided by the Wrocław Centre for Networking and Supercomputing. M.M.M. acknowledges support from the National Science Centre (Poland) under Grant No. 2024/53/B/ST3/02756. The authors thank M.~Fija\l kowski for a valuable discussion.

\bibliographystyle{elsarticle-num}
\bibliography{bibliography}

\end{document}